\documentclass[journal=paper,manuscript=article]{achemso}
\usepackage{amsmath,amssymb}

\usepackage{graphicx}
\usepackage{bm}
\usepackage{booktabs}
\usepackage{xcolor}
\usepackage[colorlinks=true,allcolors=black]{hyperref} %

\graphicspath{{figures/}}

\SectionNumbersOn %

\title{Kernel-Level Constant-Potential Molecular Dynamics in GROMACS}

\author{Sheng Bi}
\email{sheng.bi@xmu.edu.cn}
\affiliation{College of Materials, Xiamen University, Xiamen 361005, China}

\author{Mathieu Salanne}
\affiliation{Sorbonne Universit\'e, CNRS, Physicochimie des \'Electrolytes
et Nanosyst\`emes Interfaciaux, F-75005 Paris, France}
\alsoaffiliation{R\'eseau sur le Stockage Electrochimique de l'Energie (RS2E),
FR CNRS 3459, 80039 Amiens Cedex, France}
\alsoaffiliation{CNRS@CREATE, Singapore 138601, Singapore}

\author{Jun Cheng}
\affiliation{State Key Laboratory of Physical Chemistry of Solid Surfaces,
iChEM, College of Chemistry and Chemical Engineering, Xiamen University,
Xiamen, China}

\author{Guang Feng}
\email{gfeng@hust.edu.cn}
\affiliation{State Key Laboratory of Coal Combustion, School of Energy and
Power Engineering, Huazhong University of Science and Technology (HUST),
Wuhan 430074, Hubei, China}

\begin{document}
\setstretch{1.2} %

\begin{abstract}
\setstretch{1.2}
How charge rearranges within an electrode shapes the behavior of batteries,
supercapacitors, and other electrochemical devices. Molecular simulations can
capture this response by allowing the charge on each electrode atom to
adjust to its surroundings, but repeatedly calculating these charges adds
substantial computational cost. We present an open-source implementation of
the constant-potential method in GROMACS that integrates these calculations
into the program's optimized CPU and GPU routines. Comparisons with
established codes validate the electrode charges, while charging,
voltammetry, and impedance simulations
demonstrate the response to changing electrical conditions. The
implementation also supports electrodes that move or deform and models of
finite metallic screening and electron spillover. In a pair of charged
carbon nanotubes, allowing charge to move away from the facing surfaces
reduces the repulsive force at contact by about half at an intermediate
charge and keeps the pair bound where a uniform-charge model opens within
the simulated time. These capabilities make the electrical response of
electrodes accessible within GROMACS for studies of interfacial structure,
charging dynamics, and electrode motion.
\end{abstract}

\begin{tocentry}
\includegraphics[width=3.2in,height=1.7in,keepaspectratio]{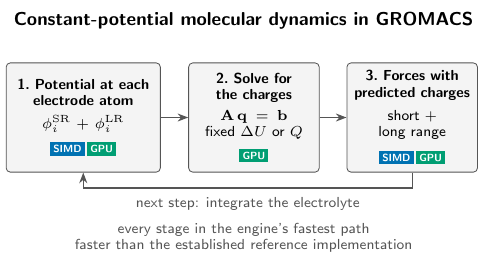}
\end{tocentry}

\section{Introduction}\label{sec:intro}

Electrode--electrolyte interfaces determine the performance of
supercapacitors, batteries, and electrocatalytic devices, and their molecular
structure, the electric double layer (EDL), is difficult to probe by
experiment alone.\cite{merlet2012,kondrat2014} Molecular dynamics reaches the
relevant nanometer and nanosecond scales, provided that the electronic
response of the electrode to the applied potential is included.
Direct quantum-mechanical treatments are costly at these length and time scales,\cite{scalfi2021,jeanmairet2022} and fixed, uniform electrode charges do not include the charge redistribution that screens the
interfacial field, which measurably changes the EDL structure, the
capacitance, and the charging dynamics.\cite{merlet2013,limmer2013} The
constant-potential method (CPM) of Siepmann and Sprik\cite{siepmann1995} and
Reed \emph{et al.}\cite{reed2007} addresses this problem: each electrode atom
is assigned a Gaussian charge whose magnitude varies so that the potential on
every electrode atom equals the imposed value, and the charges are obtained at
each timestep from a linear system whose matrix is the elastance
(inverse-capacitance) kernel of the electrode.\cite{gingrich2010} CPM is the
standard treatment beyond fixed charges,\cite{wang2014} from ionic-liquid
supercapacitors\cite{merlet2012} to nanoconfinement\cite{kondrat2014} and
electrochemical thermodynamics,\cite{limmer2013,scalfi2020} and recent reviews
summarize the state of the art.\cite{scalfi2021,jeanmairet2022,bi2024}

Established implementations exist in the dedicated program
MetalWalls,\cite{marinlafleche2020} in the ELECTRODE package of
LAMMPS,\cite{ahrensiwers2021,ahrens2022} and in ESPResSo\cite{weik2019} as
the ICC* induced-charge treatment.\cite{tyagi2010} Two obstacles to routine
use remain. First, GROMACS is a widely used MD engine,\cite{vanderspoel2005,hess2008,abraham2015}
but CPM is not a standard feature of the GROMACS~2020 release used here. Constant-potential studies with in-house GROMACS builds have been
reported.\cite{bi2020,zeng2021} Here we provide a documented implementation. Second, CPM adds per-atom electrode-potential evaluations and a charge
solve to the work of conventional MD, with an additional reciprocal-space
pass in the formulation used here. Efficient integration of this work into
the engine is therefore a practical concern. Vectorized SIMD kernels and GPU
offload,\cite{anderson2008,salomonferrer2013,pall2013,pall2020} two techniques central to the performance of modern MD, are increasingly
available for CPM: MetalWalls includes an OpenACC GPU
port,\cite{coretti2022} and OpenMM~8.4
includes a GPU-accelerated constant-potential force.\cite{openmm84cpm}
The opportunity addressed here is to integrate the electrode calculations
directly into the existing nonbonded kernels and reciprocal-space routines
of GROMACS.

In this paper we show that the constant-potential step can be integrated
by extending the kernels and PME code that a general-purpose engine already optimizes,
and we do so in GROMACS. The implementation is based on three design choices. (i) The
short-range constant-potential terms, the Gaussian--point corrections and the
per-atom electrode potential, are evaluated inside the cluster-based SIMD and
CUDA kernels of GROMACS.\cite{pall2013} The neighbor lists that GROMACS
already builds are used for the electrode--electrolyte sum, so that sum needs no separate
neighbor search. The matrix-free solver additionally maintains an electrode--electrode
neighbor list. (ii) The reciprocal-space CPM evaluations
run on the GPU: a charge pass with the electrode charges set to zero, a
per-atom potential gather from the particle-mesh Ewald (PME) grid by a
dedicated CUDA kernel, and the reciprocal-space work of every matrix-free
conjugate-gradient (CG) iteration. The dense inverse-matrix product and the
matrix-free CG iteration are also offloaded, placing the dominant
charge-solver operations on the GPU. (iii) For frozen electrodes, within
one step, the short-range and reciprocal force passes use the same set of
electrode charges, a linear extrapolation with error feedback from previous
solves. The charge pass and the solve are therefore not required before the force
evaluation, and on the GPU they overlap with the force kernel.
The formalism is unchanged; the implementation combines extensions of the
engine kernels and reciprocal-space routines with dedicated charge-solver kernels
(Sec.~\ref{sec:res-perf}). On the CUDA nonbonded path, mobile electrodes use the newly solved
charges in both force passes. The implementation also covers the protocols required in practical
electrochemistry beyond rigid electrodes at fixed
potential: fixed total charge,\cite{ahrens2022,zeng2021} single electrodes,
time-dependent $U(t)$ and $Q(t)$ protocols given as analytic expressions,
per-atom Gaussian widths, Thomas--Fermi metallicity,\cite{scalfi2020tf} electron
spillover,\cite{wang2025} flexible electrodes
through a matrix-free solver (the geometries and drives are shown in
Fig.~\ref{fig:pipeline}b), an extended-Lagrangian charge-dynamics
scheme\cite{siepmann1995} without a per-step solve, and
checkpoint restarts that preserve the full solver state. The source code is available at
\url{https://github.com/Sheng-Bi/Gromacs_CPM}.

Section~\ref{sec:theory} describes how each stage of the method is mapped
onto the SIMD and GPU code of GROMACS, Section~\ref{sec:simulations}
specifies the systems, and Section~\ref{sec:results} reports the
behavior of the solvers during charging and in the charged state, the cost relative to fixed-charge GROMACS and to CPM-LAMMPS (the ELECTRODE package of LAMMPS), the
time-dependent and Thomas--Fermi protocols, and an application to a charged nanotube bundle whose behavior depends on
electrode polarization.
\section{Theory and implementation}\label{sec:theory}

\subsection{Constant-potential formalism}\label{sec:cpm}

\begin{figure}[!htbp]
\centering
\includegraphics[width=0.98\textwidth,height=0.78\textheight,keepaspectratio]{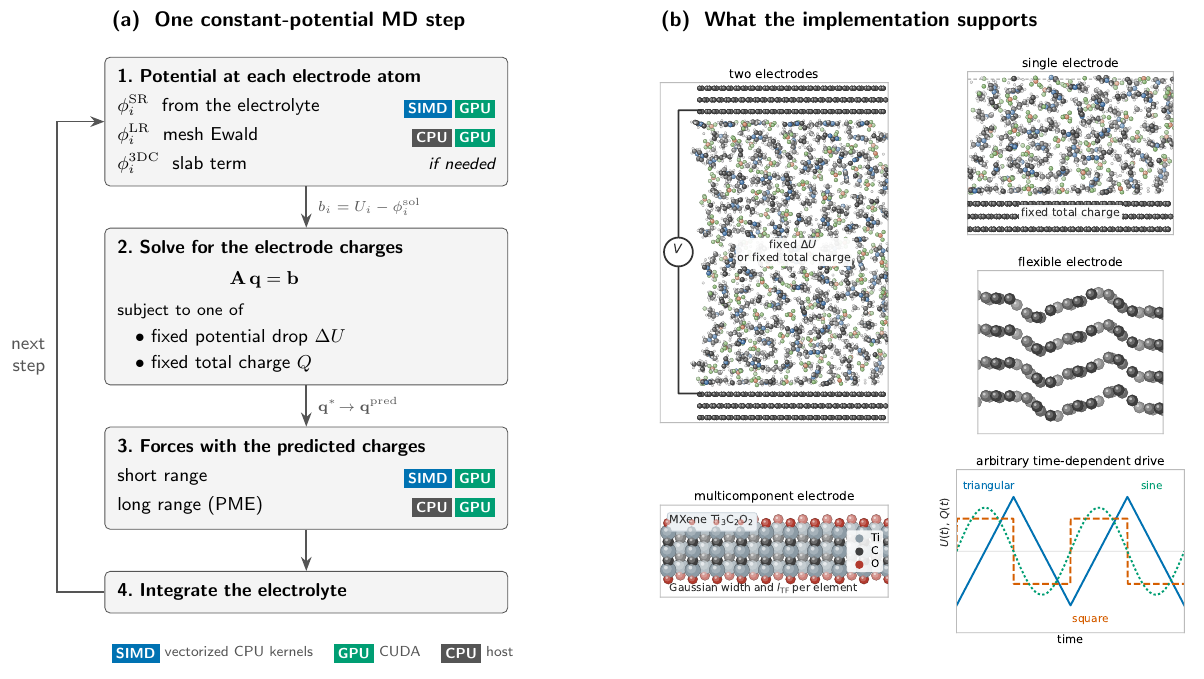}
\caption{\label{fig:pipeline}
(a) One constant-potential MD step. The potential at each electrode atom is
assembled from a short-range sum over the electrolyte, a mesh-Ewald term and,
for slab geometries, a 3DC correction. The charges follow from
$\mathbf{A}\mathbf{q}=\mathbf{b}$ under a fixed potential drop or a fixed
total charge with electroneutrality. For frozen electrodes the forces use charges predicted from previous solves,
and the new solve enters the next prediction; mobile electrodes use the
newly solved charges on the CUDA nonbonded path (Sec.~\ref{sec:impl-gpu}). Panel (a) sketches
the frozen-electrode workflow; the new solve updates the prediction for the next step. Badges mark where each stage runs:
SIMD marks the vectorized CPU kernels, GPU marks CUDA, and CPU marks the
host. With PME on the GPU, the inverse-matrix product and matrix-free CG iteration
are offloaded to the GPU. (b) Modes supported by the implementation: two electrodes under an applied voltage or a fixed
total charge, a single electrode at fixed charge, flexible electrodes,
per-element Gaussian widths on a multicomponent electrode, and an arbitrary
time-dependent drive.
}
\end{figure}

We follow the formulation of Reed \emph{et al.}\cite{reed2007} Each of the
$N_\mathrm{el}$ electrode atoms carries a Gaussian charge distribution
\begin{equation}
  \rho_i(\mathbf{r}) = q_i \left(\frac{\eta_i^2}{\pi}\right)^{3/2}
    \exp\!\left(-\eta_i^2 |\mathbf{r} - \mathbf{r}_i|^2\right),
  \label{eq:gaussian}
\end{equation}
with fluctuating charge $q_i$ and inverse width $\eta_i$. We include the
Coulomb prefactor $k_e = 1.43996$~V\,nm\,$e^{-1}$ in the kernels below,
so distances in nm and charges in $e$ give potentials in volts. The potential that
the electrode charges generate at electrode atom $i$ is
$\phi_i^\mathrm{el} = \sum_j A_{ij} q_j$, where the elastance kernel is the
sum of four terms,
\begin{equation}
  A_{ij} = A_{ij}^{\mathrm{sr}} + A_{ij}^{k} + A_{ij}^{\mathrm{self}}
         + A_{ij}^{\mathrm{3DC}}.
  \label{eq:Aij}
\end{equation}
The short-range term is the difference between the Gaussian and the
Ewald-screened point-charge interaction, evaluated within the real-space
cutoff,
\begin{equation}
  A_{ij}^{\mathrm{sr}} = k_e\frac{\operatorname{erf}(\eta_{ij} r_{ij})
                 - \operatorname{erf}(\beta r_{ij})}{r_{ij}}
  \quad (i \neq j), \qquad
  \eta_{ij} = \frac{\eta_i \eta_j}{\sqrt{\eta_i^2+\eta_j^2}},
  \label{eq:Asr}
\end{equation}
where $\beta$ is the Ewald splitting parameter. The reciprocal-space term is
the Ewald sum over the wave vectors $\mathbf{k}$ of the cell of volume $V$,
\begin{equation}
  A_{ij}^{k} = \frac{4\pi k_e}{V} \sum_{\mathbf{k} \neq 0}
    \frac{\exp(-k^2/4\beta^2)}{k^2}\,
    \cos(\mathbf{k}\cdot\mathbf{r}_{ij}),
  \label{eq:Ak}
\end{equation}
evaluated by smooth particle-mesh Ewald (PME).\cite{darden1993,essmann1995}
The self term is diagonal, $A^{\mathrm{self}}_{ii} = k_e(\sqrt{2/\pi}\,\eta_i -
2\beta/\sqrt{\pi})$. The slab term $A^{\mathrm{3DC}}_{ij} = 4\pi k_e z_i z_j/V$ is
the Yeh--Berkowitz dipole correction\cite{yeh1999} (three-dimensional Ewald
with slab correction, 3DC) for geometries with a vacuum gap; it is zero for
cells that are periodic in all three directions, such as nanotubes, which use
plain three-dimensional Ewald summation. The electrolyte charges contribute
the potential $\phi_i^\mathrm{sol} = \phi_i^\mathrm{sr} + \phi_i^{k}
+ \phi_i^\mathrm{3dc}$, given by the same three pair kernels with
$\eta_j$ taken to infinity for point charges. Every electrode atom $i$ is held at
the potential $U_i=U_i^{\mathrm{set}}+\lambda$ of its electrode,
where a common offset $\lambda$ is determined by the neutrality constraint, $\phi_i^\mathrm{el} +
\phi_i^\mathrm{sol} = U_i$, which together with electroneutrality
$\sum_i q_i = 0$ for a neutral electrolyte defines the linear system\cite{tee2022}
\begin{equation}
  \mathbf{A}\,\mathbf{q} = \mathbf{b}, \qquad
  b_i = U_i - \phi_i^\mathrm{sol},
  \label{eq:linsys}
\end{equation}
For an electrolyte of net charge $Q_\mathrm{sol}$, the corresponding
neutrality condition is $\sum_i q_i = -Q_\mathrm{sol}$. The implemented
fixed-potential mode assumes a neutral electrolyte; in charge-constraint mode
the prescribed electrode charges must balance the electrolyte charge. The system is
solved for the electrode charges at every MD step
(Fig.~\ref{fig:pipeline}a, where $\mathbf{q}^{*}$ denotes the newly solved
charge vector). In charge-constraint mode each electrode potential becomes an unknown
Lagrange multiplier, determined together with the per-atom charges by a
prescribed total $Q_\mathrm{set}$ on the designated electrode and, optionally,
an independent target on the other electrode. Two bodies with charges of the
same sign can therefore be simulated with compensating electrolyte charge. In charge-constraint
mode the resulting electrode potentials are written out every
step to obtain the voltage response of a galvanostatic protocol.

Per-atom Gaussian widths follow the MetalWalls
convention.\cite{scalfi2020,marinlafleche2020} The metallicity of the
electrode is tuned with the semiclassical Thomas--Fermi model of Scalfi
\emph{et al.},\cite{scalfi2020tf} which adds the diagonal term
$A^{\mathrm{TF}}_{ii} = 4\pi k_e l_\mathrm{TF}^2/V_i$ to Eq.~\eqref{eq:Aij}
($l_\mathrm{TF}$ is the screening length, $0$ for a perfect metal, and $V_i$
is the atomic volume). This term enters the matrix build and the matrix-free
$\mathbf{A}\mathbf{q}$ product, so it is available to every solver.
Section~S1 of the Supporting Information (SI) validates the remaining modes:
single electrodes (charge constraint only) and $U(t)$ or $Q(t)$ given as
analytic expressions that are parsed at run time.

\subsection{Short-range terms in the SIMD and CUDA kernels}\label{sec:impl-sr}

The short-range term of Eq.~\eqref{eq:Asr}, in $\mathbf{A}$ and in
$\phi^\mathrm{sol}$ alike, is the pair term
$k_e[\operatorname{erf}(\eta_{ij}r)-\operatorname{erf}(\beta r)]/r$, i.e.\ the
Gaussian--Gaussian (or Gaussian--point) interaction minus the point-charge
contribution that the reciprocal-space sum already provides. Equivalently, the
standard real-space term $k_e\operatorname{erfc}(\beta r)/r$ is replaced by
$k_e[\operatorname{erfc}(\beta r)-\operatorname{erfc}(\eta_{ij}r)]/r$. We
evaluate these terms inside the cluster-pair SIMD nonbonded kernels of
GROMACS,\cite{pall2013} which process cluster pairs in a $4\times M$ or
$2\times MM$ layout depending on the SIMD width. For every electrode atom the
kernels accumulate the short-range electrode potential $\phi^\mathrm{sr}_i$
from its electrolyte neighbors and replace the point-charge
electrode--electrolyte force by its Gaussian-screened form. Electrode atoms
are identified by a mask in the nonbonded atom data, and a per-cluster-pair
test on the masks skips the whole block unless one partner is an electrode
atom. The cluster lists are built for the short-range nonbonded interaction
in any case, so the electrode sum uses the same neighbor search and does not require a
separate list for the electrode--electrolyte sum. The same masks, widths,
pair test, and accumulation were ported to the CUDA nonbonded kernels, where
the per-atom potential is accumulated in registers and added to global memory
by atomic operations.

\subsection{A GPU-resident reciprocal-space pipeline}\label{sec:impl-gpu}

Our CPM implementation adds a reciprocal-space pass to a standard simulation: each step
needs (i) the electrolyte-only potential at every electrode site to build
$\mathbf{b}$, and (ii) a normal PME evaluation to obtain the forces. We execute both passes on the GPU:

\begin{enumerate}
\item \emph{Charge pass.} The atom charges are uploaded with all electrode
entries set to zero, spread onto the PME grid, transformed, multiplied by the
Green's function, and transformed back, all with the existing GPU-PME kernels
of GROMACS.\cite{pall2020}
\item \emph{Potential gather.} A dedicated CUDA kernel interpolates the
resulting real-space potential grid at the electrode positions with the same
$4^3$ B-spline stencils that are used for force gathering, which gives
$\phi^\mathrm{sol}_i$ for all electrode atoms in a single kernel launch.
\item \emph{Charge solve.} $\mathbf{b}$ is assembled from the SIMD/CUDA
short-range potentials, the gathered reciprocal potentials, and the 3DC
dipole term, and Eq.~\eqref{eq:linsys} is solved
(Sec.~\ref{sec:impl-solvers}).
\item \emph{Force pass.} The charges used by the short-range kernel are
uploaded, the grid is cleared, and a standard GPU-PME evaluation (spread,
FFTs, solve, force gather) produces the reciprocal forces and energy. The
newly solved charges enter the extrapolation with error feedback
$q^\mathrm{pred}(n{+}1) = 2q(n) - q(n-1) + \tfrac12\,\delta_q(n)$, with
$\delta_q(n)$ the prediction error of step $n$, which both force passes of
the next step use for frozen electrodes. For mobile electrodes on the CUDA nonbonded path the solve
precedes both force passes, which use the newly solved charges.
\end{enumerate}

The matrix-free solver needs one further electrode-only reciprocal-space evaluation of
$\mathbf{A}\mathbf{v}$ per CG iteration, where $\mathbf{v}$ is the current CG
search direction. The vector $\mathbf{v}$ is spread as the only nonzero charges, analogous to
the charge pass, and passed through the same
FFT--solve--gather sequence as stages 1--2, so the reciprocal-space work of
the iterative solver also runs on the GPU. In the mobile-electrode simulations reported here, bonded interactions
run on the CPU, and the
force pass feeds the standard force-reduction path of GROMACS, so the step
scheduling of the engine is preserved.

\subsection{Charge solvers}\label{sec:impl-solvers}

For rigid electrodes $\mathbf{A}$ is constant, and its inverse is computed
once at start-up, an $O(N_\mathrm{el}^{3})$ cost that is the linear algebra of the inversion,
not the Ewald sums, and that must be repeated
whenever the box, the Ewald parameters, or the electrode geometry changes.
Each step then solves Eq.~\eqref{eq:linsys} by one product with
$\mathbf{A}^{-1}$ followed by the electroneutrality projection. When PME runs
on the GPU the inverse stays on the device, and a pair of custom double-precision
CUDA kernels evaluates the matrix--vector product, with assembly and
constraint projection on the host. This \emph{matrix inversion} is
the default for frozen electrodes and the reference for all validation below.

For flexible electrodes $\mathbf{A}$ changes every step, and the
\emph{matrix-free CG} solver evaluates $\mathbf{A}\mathbf{v}$ during the iteration:
the reciprocal part by a PME cycle on the GPU (Sec.~\ref{sec:impl-gpu}), the
real-space part by a buffered electrode neighbor list, plus the self and 3DC
terms, with warm starts from the previous step and an electroneutrality
projection built into the iteration.\cite{ahrens2022} With PME on the GPU
the matrix-free iteration runs on the device; scalar convergence information
is returned to the host each iteration, with vector transfers at the
beginning and end of the solve. The relation between accuracy,
tolerance and cost is quantified in SI Section~S2.

Alternatively, in the \emph{extended Lagrangian}, the charges are dynamical
variables with a fictitious mass
$\mu$,\cite{car1985,siepmann1995} the Car--Parrinello-type approach of the original CPM proposal (DFT potentiostat and thermopotentiostat analogs
exist\cite{bonnet2012,deissenbeck2021}):
\begin{equation}
  \mu\,\ddot{q}_i = -(\mathbf{A}\mathbf{q}-\mathbf{b})_i ,
  \label{eq:extlag}
\end{equation}
The charges are integrated by a leapfrog step together with the atoms and,
after every step, projected onto the constraint subspace, with their
velocities projected onto its tangent. Each step then costs a single
$\mathbf{A}\mathbf{v}$ evaluation (the same device $\mathbf{A}\mathbf{q}$ as
in the matrix-free solver) instead of an iterative solve. The force pass runs
first, with the charges $q(t)$, and the charge pass afterwards, which
keeps both force passes at the same charge state. The charge
velocities start at zero, and the charges are initialized by solving
$\mathbf{A}\mathbf{q}=\mathbf{b}$ at the first step, so the dynamics starts
on the constraint manifold.

\section{Simulation details}\label{sec:simulations}

Two benchmark cells, shown as insets in Figs.~\ref{fig:val}a and
\ref{fig:perf}, represent two different regimes. The \emph{slab} is a BMIM-BF$_4$
ionic-liquid supercapacitor of 28\,944 atoms, using the CL\&P force field\cite{canongialopes2004}
with the settings specified in SI Section~S3: 736
ion pairs between two graphene electrodes of 1\,144 constant-potential atoms
each, each resting on frozen, uncharged graphite layers, with vacuum gaps
outside the electrodes treated by the 3DC correction. The
\emph{slit} is an aqueous Na$_2$SO$_4$ electrolyte in TIP4P/2005
water\cite{abascal2005} between graphite electrodes in a fully periodic cell
without vacuum gap: 82\,916 atoms, of which 4\,536 are
constant-potential atoms. All wall atoms are frozen, so
only the electrolyte is integrated. A third system, the charged nanotube
bundle, is described together with its application in
Sec.~\ref{sec:res-bundle}.

Our GROMACS simulations use the modified GROMACS~2020\cite{abraham2015}
implementation, with $\eta = 19.79$~nm$^{-1}$, PME with a
1.2~nm real-space cutoff, and a 2~fs timestep with constraints unless stated
otherwise. Cross-code comparisons use CPM-LAMMPS\cite{ahrens2022} at
matched Ewald splitting, mesh, and interpolation order. Throughput is measured on exclusive nodes of one type (two 32-core
Intel Xeon Platinum 8358 sockets, NVIDIA A100-SXM4-80GB), the CPU runs on all
64 cores and the GPU runs on one A100 with 16 host threads, as the mean of
three runs timed over 10\,000 steps after a 10\,000-step warm-up. The
conventional baseline is an unmodified GROMACS~2020 from the same
toolchain on the identical systems with static charges, in the same one-rank
configuration. Force fields, Ewald and thermostat settings, run lengths, the
benchmark protocol, and the toolchains are given in SI Section~S3.

\section{Results}\label{sec:results}

The SI presents validation checks for the implementation: the per-atom
agreement with CPM-LAMMPS (Section~S4), the consistency among the
solvers (Section~S2), and the GPU-versus-CPU agreement (Section~S5), with
charges agreeing within $4.3\times10^{-6}\,e$ rms per atom over the saved
steps of nine short validation runs.

\subsection{Charging dynamics and double-layer structure}\label{sec:res-charging}

\begin{figure}[!htbp]
\includegraphics[width=\textwidth,height=0.78\textheight,keepaspectratio]{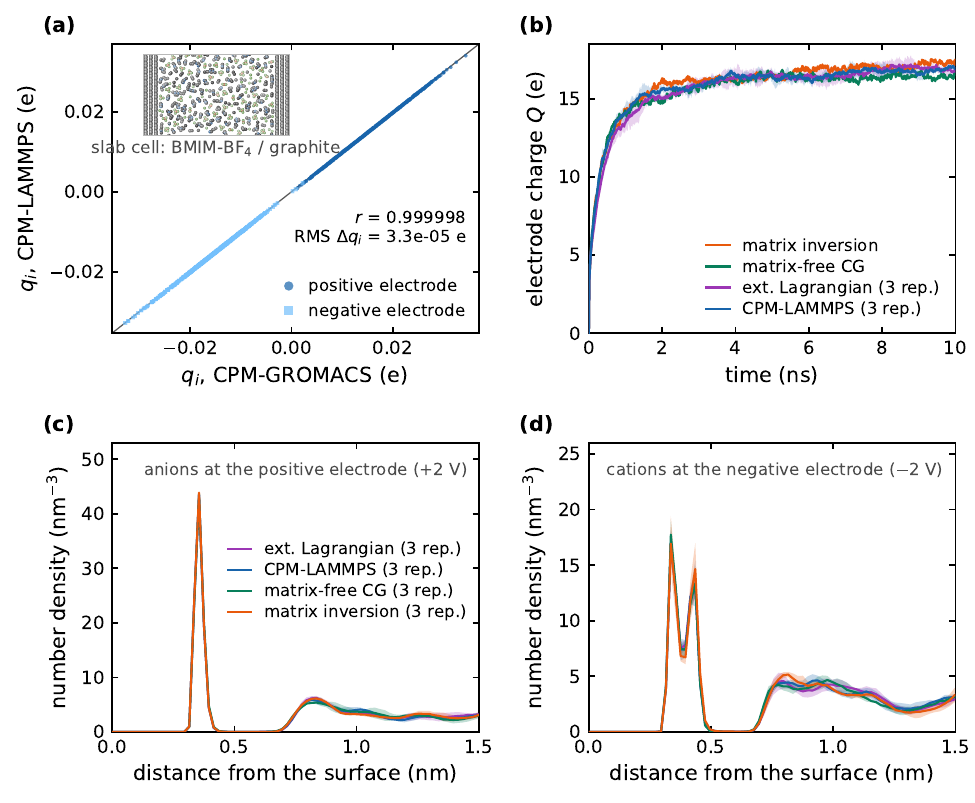}
\caption{\label{fig:val}
Validation on the slab cell, shown as an inset in (a). (a) Per-atom electrode
charges from CPM-LAMMPS against CPM-GROMACS on one equilibrated frame at
4~V. (b) Charging after a 4~V step from one unpolarized
configuration. (c) Anion density at the positive electrode and (d) cation
density at the negative electrode in the charged state, averaged over the
last nanosecond. Bands are $\pm1$ standard deviation over three replicas; the
matrix-inversion and matrix-free curves in (b) each represent one trajectory,
with the late-time averages and replica scatter reported in the text.}
\end{figure}

Figure~\ref{fig:val} compares the matrix-inversion and matrix-free solvers,
the extended Lagrangian, and CPM-LAMMPS from a single frame to the
preformed double layer. On one equilibrated frame at 4~V the per-atom
electrode charges of the two codes agree with $r = 0.999998$ and an rms
difference of $3.3\times10^{-5}\,e$, compared with a standard deviation of
the charges across atoms of $1.6\times10^{-2}\,e$ (Fig.~\ref{fig:val}a).

After a 4~V step on the unpolarized slab (Fig.~\ref{fig:val}b), all eight
trajectories charge on a similar timescale. Averaged over 5--10~ns, the
matrix-inversion and matrix-free trajectories give $16.96$ and $16.40\,e$;
the extended-Lagrangian and LAMMPS replica means are $16.69\pm0.27$ and
$16.64\pm0.14\,e$, respectively (standard deviations across three replica
means). These finite-trajectory differences are a few tenths of an electron;
the single matrix-inversion and matrix-free runs do not by themselves
establish statistical equivalence. During early charging, the
extended Lagrangian lags: over 0.45--0.55~ns it gives $11.29\pm0.28\,e$
against $11.97\,e$ for matrix inversion, consistent with a delayed response
from the fictitious charge inertia.

Figure~\ref{fig:val}c,d shows the double-layer structure of the charged
slab: the anion density at the positive electrode (c) and the cation density
at the negative electrode (d) as a function of the distance from the
electrode surface, for the four charge treatments. Every treatment starts
from the same matrix-inversion endpoint of panel (b), in three replicas that differ
only in their initial velocities, and the bands are the replica scatter. At
the positive electrode the anions form one sharp contact layer about
$0.35$~nm from the surface, followed by a depleted gap and a broad second
layer near $0.9$~nm. At the negative electrode the cation contact layer is
split into two peaks, at about $0.33$ and $0.43$~nm, followed by a second
layer near $0.9$~nm. In both panels the RMS differences between treatment-mean profiles over
the $1.5$~nm range are smaller than the typical differences between replicas
(SI Section~S6). This checks that the treatments maintain the same
preformed double layer; the common starting structure and slow ion exchange
do not establish independent equilibration.

\subsection{Performance}\label{sec:res-perf}

\begin{figure}[!htbp]
\centering
\includegraphics[width=\textwidth,height=0.78\textheight,keepaspectratio]{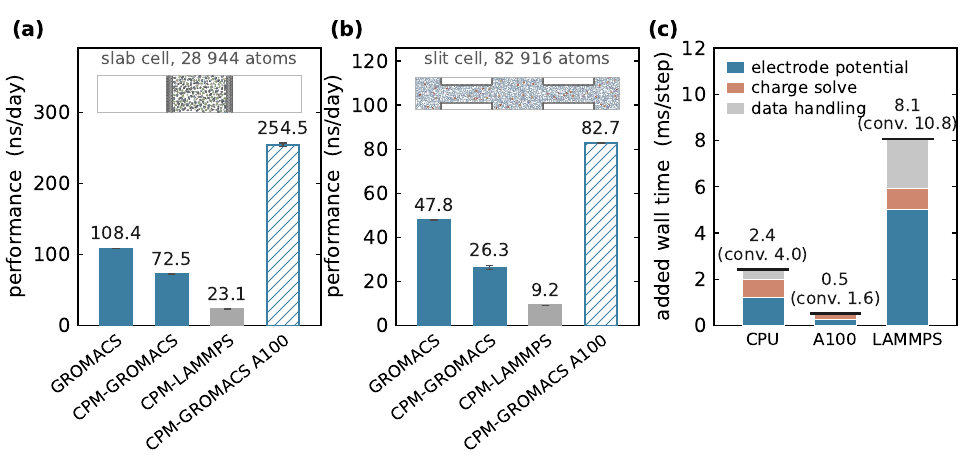}
\caption{\label{fig:perf}
Throughput of conventional GROMACS, CPM-GROMACS, and CPM-LAMMPS for (a) the
slab cell (28\,944 atoms) and (b) the slit cell (82\,916 atoms), each shown
as an inset. Bars are the mean of three 10\,000-step timings taken in series
on exclusive nodes of the same type, GROMACS as one rank with 64 threads and LAMMPS
as 64 MPI ranks; the hatched bar is CPM-GROMACS on one A100 with
16 host threads. (c) What constant potential adds to each
step of the slit cell, measured by counters inside the code, for
CPM-GROMACS on the node (CPU), CPM-GROMACS on the A100, and CPM-LAMMPS, with
the conventional step time of each row in parentheses; the segments sum to the
net, marked by the tick. Error bars in (a,b) are sample standard deviations.
The data-handling segment in (c) includes the residual needed to reconcile
the component counters with the measured net time.}
\end{figure}

Figure~\ref{fig:perf} shows the cost of constant potential relative to a conventional simulation of
the same system, and how this compares with CPM-LAMMPS (the reference
code below), using the CPU configurations described in SI Section~S7. On the
slab cell (Fig.~\ref{fig:perf}a; 28\,944 atoms) conventional GROMACS, the unmodified
package with fixed electrode charges, reaches $108.4\pm0.5$~ns/day and
constant potential $72.5\pm0.1$; the reference code
reaches $23.1\pm0.03$~ns/day, so CPM-GROMACS runs $3.1$ times faster. On
the slit cell (Fig.~\ref{fig:perf}b; 82\,916 atoms) the same protocol gives
$47.8\pm0.2$~ns/day conventional against $26.3\pm0.9$ constant-potential
for GROMACS, and $9.21\pm0.02$ for the reference code,
a throughput ratio of $2.9$.

Figure~\ref{fig:perf}c shows the origin of the added cost, using counters
inside both codes. In our implementation the electrode sum adds no neighbor-search
cost, because the cluster lists are already
built and the Gaussian terms are evaluated inside the kernels. The remaining measured cost comes from the electrode calculations and data
handling: evaluating the electrode potentials costs
$1.23$~ms per step ($0.58$ of it in the nonbonded kernel), the charge solve
$0.78$ and the residual attributed to data handling $0.41$, a net $2.41$~ms on top of the $3.99$~ms
conventional step. In the reference code the same three items cost $5.03$,
$0.91$ and $2.13$~ms, $8.07$ in total; in that code the potential evaluation is performed by the
ELECTRODE fix, which builds and loops over its own neighbor lists.

The relative cost differs on the GPU. The benchmarked ELECTRODE implementation runs on CPUs, so the hatched bars of Fig.~\ref{fig:perf} compare our GPU
path with our own CPU runs. One A100 with 16 host threads, with the inverse-matrix product
on the device, runs constant potential at $254.5\pm2.7$~ns/day on the
slab and $82.7\pm0.1$ on the slit, $3.5$ and $3.1$ times the 64-core node.
In the breakdown the electrolyte-only reciprocal pass costs $0.25$~ms per
step, the device solve $0.24$ and the net data-handling residual $0.03$, so the
constant-potential step costs $0.52$~ms more than the $1.56$~ms conventional
step.

\subsection{Time-dependent protocols and impedance spectroscopy}\label{sec:res-apps}

\begin{figure}[!htbp]
\includegraphics[width=\textwidth,height=0.78\textheight,keepaspectratio]{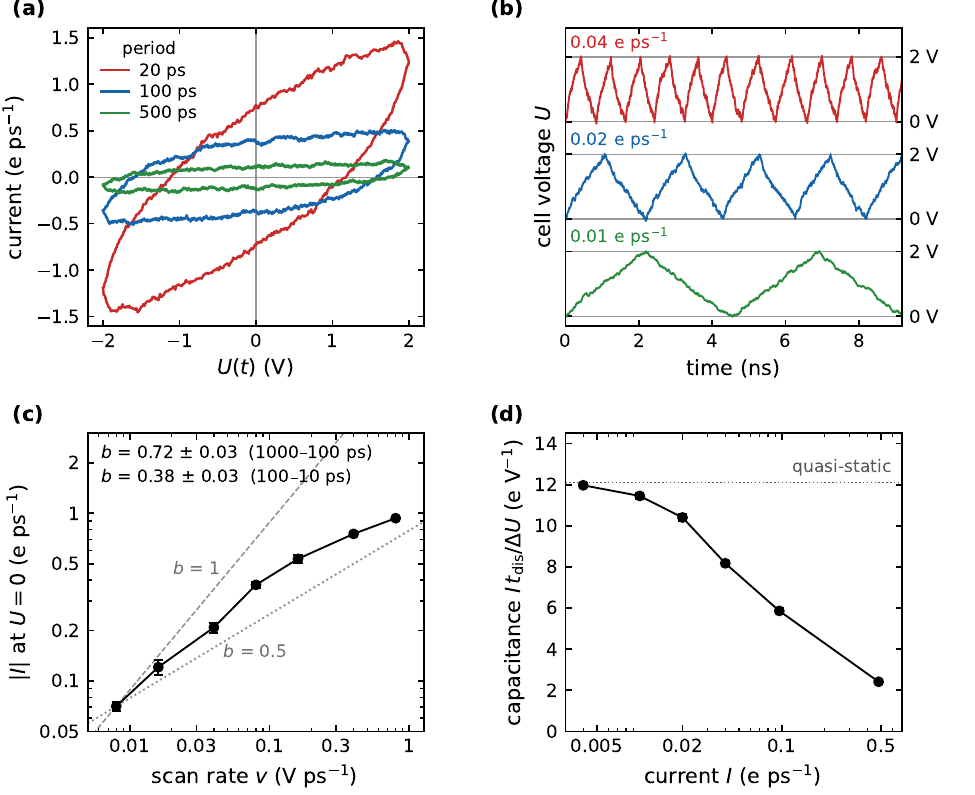}
\caption{\label{fig:apps}
Time-dependent protocols in the slit cell, cycle-averaged; the protocols and
further analyses are in SI Sections~S3 and S8. (a) Cyclic voltammetry at
periods of 20~ps (red), 100~ps (blue) and 500~ps (green): the slowest sweep
gives the approximately rectangular voltammogram of a capacitor, the fastest is
kinetically limited. (b) Galvanostatic cycling between 0 and 2~V at
$I = 0.04$ (red), $0.02$ (blue) and $0.01\,e$/ps (green), as the cycler reads
the voltage; the traces are offset vertically and gray lines mark the cut-offs
of each. (c) Current at $U = 0$ against the scan rate for seven sweeps with
periods from 10 to 1000~ps, with $b$ fitted above and below 100~ps; the guide
lines have slopes $b = 1$ (capacitive) and $0.5$ (transport-limited).
(d) Capacitance from the discharge time against the galvanostatic current,
with the quasi-static value dotted. Error bars in (c) and (d) are standard
errors over cycles.
}
\end{figure}

Electrochemical capacitance and kinetics can be probed by sweeping, stepping
or cycling the bias.\cite{zeng2021,jeanmairet2022} We therefore test
time-dependent potential and charge protocols.
We start from a slit cell equilibrated at 0~V; its potential-step discharge
gives a charge-relaxation time $\tau$ of about $0.15$--$0.2$~ns. Symmetric
triangular potential sweeps between $-2$ and $+2$~V with periods of 20, 100
and 500~ps span half-cycle durations below and above this relaxation time, and the
simulation reproduces finite-scan-rate voltammetry (Fig.~\ref{fig:apps}a;
after 1~ns of cycling every later cycle is averaged, 250, 50 and 10 cycles).
The 500~ps sweep gives the approximately rectangular voltammogram of a capacitor,
with a mean branch-current magnitude of $0.11\,e$/ps over $|U| < 1$~V, from which
the differential capacitance $C = I/(\mathrm{d}U/\mathrm{d}t)$ is
$6.8\,e$/V. At 100~ps the mean branch-current magnitude is $0.37\,e$/ps, but
the capacitance has decreased to $4.6\,e$/V. At 20~ps the branches slope
and change sign, and the apparent capacitance is $1.8\,e$/V. A shorter half-cycle forms less of the
double layer, and at 20~ps the double layer is still relaxing from the
previous half-cycle when the next one begins, which produces the sloping
branches.

In charge-constraint mode the cell can also be cycled as in a galvanostatic
experiment. A constant current $\pm I$ charges the cell until the voltage
reaches 2~V and discharges it until the voltage falls to 0~V. The current is
reversed when a moving average of the realized cell voltage reaches a
cut-off. Figure~\ref{fig:apps}b shows the voltage over the same 9.2~ns at
$I = 0.04$, $0.02$ and $0.01\,e$/ps, which complete 11, 4 and 2 cycles of
818~ps, 2.08~ns and 4.58~ns. The protocol works as in an experiment: in
every cycle the current reverses at both cut-offs, the voltage changes almost
linearly between them without a step at the reversals, and the charge
returned on discharge equals the charge stored on charging within the
cycle-to-cycle scatter. The lower the current, the closer the cycle comes to
the ideal triangle of a capacitor: the capacitance from the slope of the
discharge branch approaches the one from the discharge time (SI Table~S3),
and at equal electrode charge the charging branch lies above the discharging
one by 0.51, 0.34 and 0.18~V at 0.04, 0.02 and $0.01\,e$/ps.

\begin{figure}[!htbp]
\includegraphics[width=\textwidth,height=0.78\textheight,keepaspectratio]{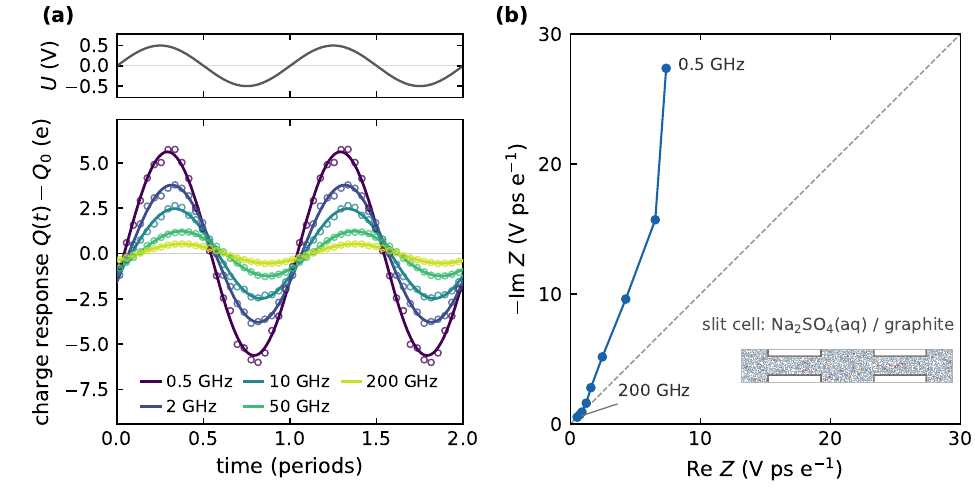}
\caption{\label{fig:eis}
Impedance spectroscopy in the slit cell, shown as an inset: sinusoidal drives
of amplitude 0.5~V about 0~V at nine frequencies from 0.5 to 200~GHz. (a) The
drive $U(t)$ (upper) and the charge response at five of the frequencies
(lower), against time in units of the period; circles are the charge folded
over every whole cycle after the first, lines are the fitted sinusoids.
(b) Nyquist plot of $Z = \hat{U}/\hat{I}$; the dashed line is the Warburg
response that the high-frequency points follow. The capacitances $C'$ and
$C''$ quoted in the text come from the same fits.
}
\end{figure}

These features are consistent with a delayed ionic response in the pore.\cite{zeng2021} The ions have to move
along the slit to charge the walls, so the double layer lags the imposed
charge, and the charge left at a cut-off is roughly the current times this
lag. A half-cycle of 1--2~ns at the two lower currents is several
charge-relaxation times long, and the cell approaches capacitive behavior. A distributed pore resistance provides a plausible interpretation of this
hysteresis. No lumped series resistor is imposed in the model, and the analysed
traces do not resolve a distinct instantaneous voltage step at reversal. The gap defines an effective
resistance $R = \Delta U/2I$ of 6.4, 8.6 and 9.0~V\,ps\,$e^{-1}$, which
is comparable in magnitude to the low-frequency impedance reported below
(Fig.~\ref{fig:eis}). Faster currents, with half-cycles
down to 10~ps, are analysed in SI Section~S8.

Figures~\ref{fig:apps}c and d give the standard analyses of the two
protocols (further quantities in SI Section~S8). For the voltammograms, the
current at fixed potential grows with the scan rate as $I \propto v^{b}$,
with $b = 0.72 \pm 0.03$ over the sweeps with periods from 1000 to 100~ps
and $0.38 \pm 0.03$ from 100 to 10~ps at $U = 0$, and similar values at $\pm1$~V. For the galvanostatic
cycles, the capacitance from the discharge time,
$C = I\,t_\mathrm{dis}/\Delta U$, is close to the quasi-static value at the
lowest currents, 12.0 and 11.5~$e$/V at 0.004 and $0.01\,e$/ps, and falls
to 10.4, 8.2, 5.9 and 2.4~$e$/V at 0.02, 0.04, 0.096 and $0.48\,e$/ps. In
experiments, $b = 0.5$ usually points to diffusion-controlled faradaic
charging. Here there are no faradaic reactions: $b$ decreases because the
ions cannot move along the slit fast enough to follow the faster sweeps, the
same transport limitation that lowers the galvanostatic capacitance at high
current.

The same controller drives a sinusoidal bias $U(t) = U_0\sin(2\pi f t)$ with
$U_0 = 0.5$~V about 0~V at nine frequencies from 0.5 to 200~GHz, each an independent run from the same 0~V frame. Fitting the charge response after
the first cycle gives the complex capacitance $C(f) = C' - iC''$ and the
impedance $Z = U_0/(2\pi i f\,\hat{Q})$ (Fig.~\ref{fig:eis}a). At 0.5~GHz
the charge follows the potential with an amplitude of $5.6\,e$ and a lag of
$15^\circ$; at 200~GHz the amplitude has decreased to $0.53\,e$ and the lag
is $44^\circ$. From 50 to 200~GHz the Nyquist points (Fig.~\ref{fig:eis}b) lie close to a
$45^\circ$ line, consistent with transport-limited (Warburg) charging.
Between 1 and 5~GHz the phase of $Z$ is approximately $-65$ to
$-67^\circ$, while at 0.5~GHz it reaches $-75^\circ$, moving toward the
$-90^\circ$ limit of a capacitor. $C'$ increases from $0.76\,e$/V at 200~GHz to $10.8\,e$/V at
0.5~GHz, approaching the quasi-static $Q(2\,\mathrm{V})/2\,\mathrm{V} =
12.1\,e$/V, and the largest sampled $C''$ occurs at 1~GHz, where the potential-step relaxation
time predicts $1/2\pi\tau$ of about $0.8$--$1$~GHz. These features are consistent with ion-transport-limited charging in the slit. The ions cannot follow the faster
drive, so the amplitude falls and the lag grows with frequency; at high
frequency the current is limited by the transport along the channel, which
gives the Warburg line; and the lower-frequency response reflects progressively more complete
charging of the pore. The differential capacitances of the three
voltammograms, $6.8$, $4.6$ and $1.8\,e$/V at periods of 500, 100 and
20~ps, are close to $C'(f)$ at $f = 1/T$, $6.9$, $4.3$ and $1.8\,e$/V, and so do
the galvanostatic capacitances at 0.04 and $0.096\,e$/ps, 8.2 and
5.9~$e$/V against $C'$ of 8.1 and $5.4\,e$/V interpolated at their cycle
frequencies of 1.2 and 4.1~GHz. The step, sweep and sinusoidal protocols thus give a consistent dynamical
response of the cell, although the drive amplitudes and mean biases differ,
so the comparison is qualitative rather than an equality of response functions.

\subsection{Thomas--Fermi metallicity and electron spillover}\label{sec:res-tf}

\begin{figure}[!htbp]
\includegraphics[width=\textwidth,height=0.78\textheight,keepaspectratio]{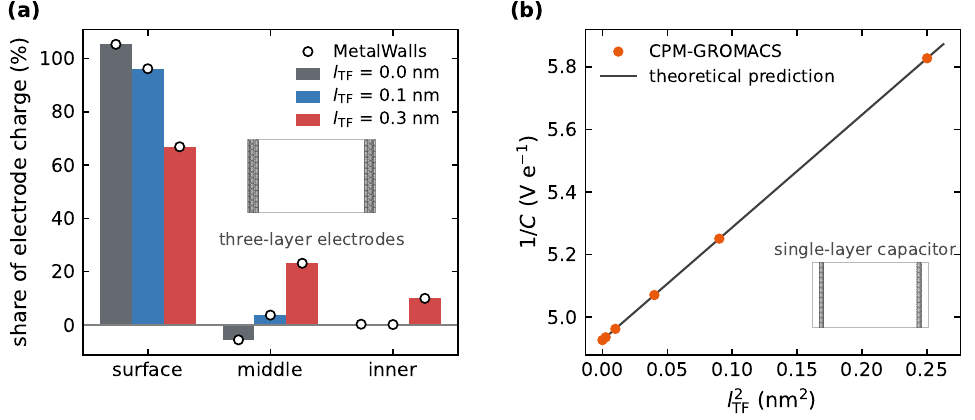}
\caption{\label{fig:tf}
Thomas--Fermi metallicity. (a) Share of the electrode charge on each layer of
the three-layer graphene electrode shown as an inset, at 4~V for
$l_\mathrm{TF} = 0$, $0.1$ and $0.3$~nm (bars, CPM-GROMACS; open circles,
MetalWalls on identical coordinates). The perfect metal overscreens its
surface, so the middle layer is negative; a finite screening length lets the
charge penetrate. (b) Inverse capacitance of the single-layer capacitor shown
as an inset against $l_\mathrm{TF}^2$, with the analytical prediction quoted
in the text. The two panels use different cells because penetration requires
more than one layer, while the analytical result holds only for one.
}
\end{figure}

Constant-potential simulations usually treat the electrode as a perfect
conductor, which confines the induced charge to its
surface.\cite{siepmann1995,reed2007} Real electrode materials screen the
field over a finite length, so the charge spreads into the electrode and the
capacitance decreases; the semiclassical Thomas--Fermi model of
Section~\ref{sec:cpm} describes this with a single screening length
$l_\mathrm{TF}$.\cite{scalfi2020tf}
Figure~\ref{fig:tf} shows the model applied to a three-layer graphene
capacitor with vacuum between the electrodes (parameters in SI
Section~S3). For the perfect
metal ($l_\mathrm{TF}=0$) the surface layer carries $105\%$ of the electrode
charge and the layer below it $-6\%$, i.e.\ a metal overscreens its own
surface. With $l_\mathrm{TF} = 0.1$ and $0.3$~nm the charge extends into the
electrode: the
surface share decreases to $96\%$ and $67\%$, at the larger length the middle
and inner layers carry $23\%$ and $10\%$, and the capacitance decreases.
We ran MetalWalls\cite{marinlafleche2020} on the same coordinates, and the
two codes agree atom by atom, to at most $6\times10^{-7}\,e$ rms, at all
three screening lengths (Fig.~\ref{fig:tf}a). For a single-layer capacitor the model is analytical:
each electrode adds $4\pi k_e l_\mathrm{TF}^2/(N V_i)$ to its inverse
capacitance, so $1/C$ is linear in $l_\mathrm{TF}^2$ with slope
$2k_e\cdot4\pi/(NV_i) = 3.603$~V\,$e^{-1}$\,nm$^{-2}$, where
$k_e = 1.440$~V\,nm\,$e^{-1}$ is the Coulomb constant and $N = 1\,144$ is
the number of atoms per electrode; the measured slope is $3.603$
(Fig.~\ref{fig:tf}b). Together, the two tests validate the implementation of
the Thomas--Fermi model.

The code also supports the electron-spillover model of Wang
\emph{et al.} (sCPM), in which the Gaussian charges of the outermost
electrode layer sit a distance $r_D$ outside the nuclei, at the center of
the induced electron density, which the original study found to bring the computed capacitance of
Au(111) into agreement with experiment.\cite{wang2025} For rigid electrodes
it needs no change to the code: the displaced charge sites are the electrode
atoms, and the nuclei are kept as uncharged atoms that carry the
Lennard-Jones interaction. Moving both charge planes of a single-layer
capacitor inward by $r_D$ lowers $1/C$ with a fitted-to-analytical slope ratio of $1.00001$ for matrix inversion
(analytical slope $-2k_e\cdot4\pi/A$), and a three-layer capacitor agrees
with MetalWalls on the same coordinates to $4\times10^{-7}\,e$ rms per atom.
For a reconstructed cell based on the original study, with Au(111) electrodes 10~nm apart in 2~M
aqueous NaClO$_4$, the implementation also reproduces the effect the model
was built for: the capacitance per electrode rises from $7.8\pm0.2$ to
$46\pm1\,\mu$F\,cm$^{-2}$ when the charges are displaced. This reproduces
the enhancement, but exceeds the approximately 6 and 24--41 reported for the
two treatments.\cite{wang2025} These finite-time integral estimates are
not a quantitative reproduction of the published differential-capacitance
curve: the reconstructed electrolyte parameters and residual interfacial
relaxation limit the comparison (SI Section~S1).

\subsection{Flexible electrodes: charging of a nanotube bundle}\label{sec:res-bundle}

\begin{figure}[!htbp]
\includegraphics[width=\textwidth,height=0.78\textheight,keepaspectratio]{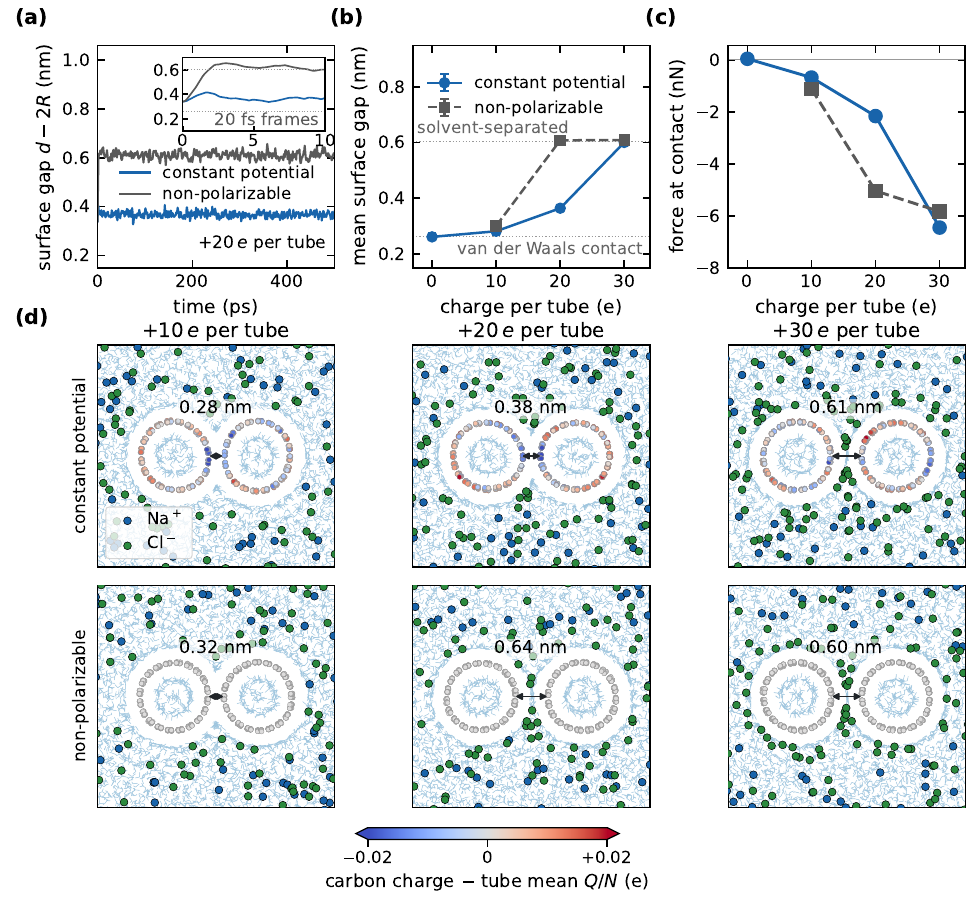}
\caption{\label{fig:bundle}
A two-nanotube bundle in 1~M NaCl with self-consistent charges (blue) or
uniform charges $q_i=Q/N$ (gray), at the same total charge per tube.
(a) Surface gap $d-2R$ at $+20\,e$ over 500~ps; inset: the first 10~ps at
20~fs resolution. (b) Mean gap over the last 200~ps of each run, with temporal
standard deviations. Dotted lines mark contact and solvent-separated states.
(c) Constraint force at the common contact distance $d_0=1.6172$~nm,
negative for repulsion; means over 20--120~ps at $+10$ and $+20\,e$,
200--500~ps at $+30\,e$, and 5--100~ps for the neutral reference.
Error bars are standard errors from ten time blocks. The neutral points in
(b,c) are self-consistent references; no neutral uniform-charge trajectory is
plotted. (d) Final end views,
self-consistent above uniform charges, with carbons colored by $q_i-Q/N$.
Na$^+$ is blue, Cl$^-$ green, water is shown as bonds, and arrows span the
surface gaps.
}
\end{figure}

Many electrode processes change the shape of the electrode itself: ions
intercalate between the layers of two-dimensional materials, and nanotubes
bundle or separate as they are charged. Simulating them requires a
constant-potential method with mobile electrode atoms, which the matrix-free
solver provides (Section~\ref{sec:impl-solvers}); we demonstrate this feature
on the charging of a nanotube bundle. The two tubes are free to move, so
their spacing is set by the balance of forces at their surfaces, and the
question is whether a self-consistent electrode charge changes that
balance. The tubes are armchair (10,10) single-wall nanotubes of $N = 800$
carbons, bonded across the periodic boundary so that they are effectively
infinite, in TIP3P water with 1~M NaCl, run with the matrix-free solver
under the charge constraint with an independent target on each tube
(SI Sections~S3 and S9). Two tubes in van der
Waals contact are held at $+Q$ each and compared with the same systems on a
non-polarizable electrode that carries $Q/N$ on every carbon
(Fig.~\ref{fig:bundle}; one trajectory per charge and treatment).

The difference between the two treatments starts at the charge
distribution (Fig.~\ref{fig:bundle}d). When the tubes are in contact under
constant potential, the charge leaves the facing sides and accumulates on
the far sides: at $+20\,e$ per tube the facing carbons carry
$13\times10^{-3}\,e$ each and the outer carbons $27\times10^{-3}$, averaged
over the 1~ns trajectory for atoms in the facing and outer $60^\circ$ sectors
of the final geometry, whereas
the non-polarizable model places $25\times10^{-3}$ on every carbon. This redistribution is consistent with reduced repulsion across the contact;
the constrained-force and free-trajectory comparisons below quantify the
change in total force and opening behavior.

The gap between the tubes shows the consequence (Fig.~\ref{fig:bundle}a,b).
At $+10\,e$ both treatments stay in contact, with surface gaps of
$0.28$ and $0.30$~nm. At $+20\,e$ ($0.15$~C/m$^2$) the constant-potential
pair remains in close contact throughout the 1~ns run, with a mean inter-axis
distance of $1.72$~nm and a surface gap of $0.36$~nm over the last 200~ps,
while the non-polarizable pair
opens within the first picoseconds to the solvent-separated state at
$0.61$~nm. At $+30\,e$ water and ions enter the gap in both treatments: the
tubes separate within 100~ps into a solvent-separated pair at $1.96$~nm,
with averages of $10.3$ and $6.8$ Cl$^-$ within $0.45$~nm of both tubes
for self-consistent and uniform charges, respectively (ten frames over
the final 100~ps). Both models therefore reach a solvent-separated geometry,
although their ion populations differ.

The force at contact (Fig.~\ref{fig:bundle}c), measured with the same
constrained protocol for both treatments, quantifies the same effect.
Because its charge cannot redistribute, the non-polarizable model
gives stronger repulsion between the tubes in contact: $1.6$ times at
$+10\,e$ and $2.3$ times at $+20\,e$ ($-1.12$ and $-5.05$~nN against $-0.69$
and $-2.17$~nN). This is consistent with the opening observed for the
non-polarizable bundle at $+20\,e$. At $+30\,e$ the forces at the same
contact distance are $-6.46$ and $-5.84$~nN for self-consistent and uniform
charges, respectively: uniform charges give about $10\%$ less repulsion.
Thus the large intermediate-charge difference is reduced and its sign reverses
at the highest sampled charge. These constrained forces do not measure the
force in the solvent-separated state or a free-energy barrier to opening
(SI Table~S4).

A fixed uniform charge thus reproduces the two end states, the bound bundle
at low charge and the solvent-separated pair at high charge, but gives a
different outcome at $+20\,e$: it predicts about twice the repulsive force
at contact and opens the bundle during the simulated trajectory. With one
trajectory per treatment and charges spaced by $10\,e$, these observations
show a polarization-dependent opening tendency, rather than a precisely
determined equilibrium separation threshold.

\section{Conclusions}\label{sec:conclusions}

The implementation integrates self-consistent electrode response into the
CPU and GPU execution paths of GROMACS. With the Gaussian-charge short-range terms in the SIMD
and CUDA nonbonded kernels, with the reciprocal-space evaluations and dominant
charge-solver operations offloaded to the GPU, constant-potential dynamics costs $1.50$--$1.82$ times a
conventional fixed-charge simulation of the same system on CPUs and runs
$2.9$--$3.1$ times faster than CPM-LAMMPS at matched Ewald settings, and a
single A100 GPU accelerates it by a further factor of $3.1$--$3.5$ over a
64-core node. The implementation is validated per atom against that reference code and
across its own solvers and GPU path, and covers the modes that practical
electrochemistry requires: cyclic voltammetry,
galvanostatic cycling with voltage cut-offs and impedance spectroscopy give
consistent capacitances at comparable frequencies, and the Thomas--Fermi
and electron-spillover models agree atom by atom with MetalWalls. Flexible electrodes under an independent charge constraint
give the main physical result of this paper: a metallic nanotube pair at prescribed total charge on each tube
redistributes its charge away from the contact and
has about half the contact repulsion of the uniform-charge model at
$+20\,e$ per tube. The simulated pair remains bound at this charge and
opens at $+30\,e$, whereas the uniform-charge pair opens already at $+20\,e$;
the sampled charges and trajectory lengths do not locate an equilibrium
opening threshold. The implementation currently supports a single MPI rank;
configuration requirements are provided in the repository documentation.
Mobile-electrode production simulations require the CUDA nonbonded path to
use freshly solved charges in both force passes; the CPU path retains
extrapolated charges and is not validated here for production dynamics. We expect the implementation to enable the routine use of constant-potential
simulations in GROMACS.

\begin{suppinfo}
Validation of the remaining features, simulation details, and further
analyses (PDF).
\end{suppinfo}

\begin{acknowledgement}
This work was supported by the New Generation Artificial
Intelligence---National Science and Technology Major Project
(2025ZD0122703), the National Natural Science Foundation of China
(Grant Nos.~22503072, T2325012, and 22225302), and the Fundamental Research Funds for the Central
Universities (20720250122).
\end{acknowledgement}

\section*{Data Availability Statement}
The modified GROMACS source, example inputs, and analysis scripts are
available at \url{https://github.com/Sheng-Bi/Gromacs_CPM}
(branch \texttt{main}). A versioned archive of the code
and supporting simulation data, with a DOI, will accompany publication.

\bibliography{references_v2}

\end{document}